\documentclass{article}

\usepackage{amsmath,amsthm,amssymb,mathrsfs,bbm}
\usepackage{accents}

\usepackage{color,soul}
\usepackage{graphicx}
\usepackage[hidelinks]{hyperref}
\usepackage{booktabs}
\usepackage{bookmark} 

\usepackage[toc,page]{appendix} 

\usepackage{natbib} 
\setcitestyle{aysep={}} 

\usepackage{ragged2e} 

\usepackage[left=1.25in,right=1.25in, top=1.25in, bottom=1.25in, footskip=.25in]{geometry} 
\usepackage{setspace}\setdisplayskipstretch{} 
\title{Some Economics of Artificial Superintelligence\thanks{hathomp@olemiss.edu. Department of Economics, University of Mississippi, Odom Hall, University, MS 38677. I thank Peter Leeson, Cullen O'Keefe, and Philip Trammell for thoughtful comments and suggestions. This project would not be possible without generous support from the Institute for Humane Studies.}}
\author{Henry A. Thompson}
  \vspace{7mm}

\begin{document}

\maketitle
\begin{abstract}
Conventional wisdom holds that a misaligned artificial superintelligence (ASI) will destroy humanity. But the problem of constraining a powerful agent is not new. I apply classic economic logic of interjurisdictional competition, all-encompassing interest, and trading on credit to the threat of misaligned ASI. Even while granting AI-safety canon some of its strongest assumptions, I show that an acquisitive ASI refrains from full predation under surprisingly weak conditions. When humans can flee to rivals, inter-ASI competition creates a market that tempers predation. When trapped by a monopolist ASI, its ``encompassing interest" in humanity's output makes it a rational autocrat rather than a ravager. And when the ASI has no long-term stake, our ability to withhold future output incentivizes it to trade on credit rather than steal. In each extension, humanity's welfare progressively worsens. But each case suggests that catastrophe is not a foregone conclusion. The dismal science, ironically, offers an optimistic take on our superintelligent future.
\end{abstract}

\vspace{0.25in}


\newpage

\section{Introduction}
Human history is replete with examples of the strong preying upon the weak. Viking longships raided vulnerable areas of coastal Europe. Feudal lords extracted tribute from their much weaker serfs. Leviathan governments past and present have routinely taken from their far weaker subjects and neighbors. It is no surprise then to hear political philosophers like Montesquieu concede that ``constant experience shows us that every man invested with power is apt to abuse it, and to carry his authority as far as it will go" \citep[197]{montesquieu1777}.

Conventional wisdom says that artificial superintelligence (ASI) will be no different. A computer system with alien motives and abilities that far exceed the best humans would have no interest in cooperating. It would simply take what it wants without regard for the welfare of humans. The result? Existential catastrophe \citep{bostrom2014, dung2023current, carlsmith2023existential}.

I argue that catastrophe caused by misaligned and acquisitive ASI is not a foregone conclusion. Rather, ASI's treatment of humans depends upon a familiar logic: the incentives it faces. To do so, I present a simple, benchmark model in which a misaligned and acquisitive ASI decides whether to trade with or steal from humans. The ASI cooperates when humanity can credibly impose sanctions for misbehavior. 

I proceed by progressively weakening humanity's position, thereby granting AI-safety canon some of its strongest assumptions: that ASI is misaligned, acquisitive, and stronger than humans. I turn first to a world in which the ASI is too powerful to be sanctioned but must compete with rivals for humanity's output. As humans can move between rivals, there is a ``market for theft" that checks the local ASI's incentive for unrestrained predation.\footnote{The logic of this extension shares much in common with \citet{tiebout1956pure}.} The reason is that unrestrained predation causes humans to flee, depressing its future income. To the great benefit of humanity the local ASI tempers its takings, thereby avoiding catastrophe. 

Then I turn to a world in which ASI is so strong that humans are trapped. They have no exit option. Despite being trapped by an extraordinarily strong ASI, the extension shows that humanity avoids catastrophe as long as the ASI is sufficiently patient.\footnote{This extension draws on Olson's \citeyearpar{olson1993dictatorship} and Mcguire and Olson's \citeyearpar{mcguire1996economics} models of predatory government.} The reason is that the ASI's strength gives it an ``encompassing interest" in the affairs of humans. The less productive humanity is, the less the ASI has to take. As a result even a paperclip maximizing ASI is incentivized to rule like a rational autocrat. It takes some but not all of what humans make. Taking more risks depressing its future takings.

I conclude with the bleakest scenario yet: a powerful ASI with no long-term stake in humanity's future. Here, predation seems inevitable. But the extension shows that humanity holds its future production hostage.\footnote{This extension draws heavily on Leeson's \citeyearpar{leeson2007trading} model of banditry in late precolonial west central Africa.} An ASI cannot steal that which does not yet exist. By threatening to produce only for subsistence, humans make predation unprofitable. This forces a patient ASI to pay in advance for the very goods it would otherwise steal. Trading on credit enriches both humanity and the ASI. Humanity avoids subsistence living and the ASI avoids looking elsewhere for the resources it needs.

In each case, humanity's welfare progressively worsens. But each case suggests that catastrophe is not a foregone conclusion. Indeed the model and its extensions suggest that, for the very worst outcome to occur, three conditions must hold simultaneously. First, humanity cannot impose credible sanctions. Second, humans have no exit option. Third, ASI must be myopic. Should all three conditions be present, humanity does not do well. Otherwise, it fares better. As a result the standard narrative of ``power $+$ misalignment $=$ doom" is incomplete. The classic logic of \citet{tiebout1956pure, olson1993dictatorship}, and \citet{leeson2007trading} suggests that the conditions for catastrophe are narrower still.

The problem of constraining an agent far stronger than its subjects is not new. It has long plagued economists and political scientists alike \citep{buchanan1975limits, brennan1980power, north1981structure, north1989constitutions, weingast1993constitutions, weingast1997political, greif1994coordination, barzel2000property, barzel2002theory, bates2002organizing, acemoglu2006economic, leeson2007arrgh, leeson2007anarchy}. Such work suggests that strong agents are more likely to act in the interests of weaker ones when, for instance, there is interjurisdictional competition \citep{weingast1995economic, qian1997federalism}, the strong have stable and encompassing interests \citep{barzel1992confiscation, olson1993dictatorship, mcguire1996economics}, and the weak can trade on credit \citep{leeson2007trading}.\footnote{A related body of work is that which concerns private governance. Without a third-party enforcer to constrain AI misbehavior, humans and AI are in a world of statelessness. But economists have shown that cooperation is robust to statelessness in a huge variety of contexts. See, for example, \citet{demsetz1967toward, greif1989reputation, ellickson1991order, bernstein1992opting, anderson1994raid, leeson2009laws, leeson2014anarchy, thompson2024industrial}.} None have considered whether the same logic also applies to alien intelligences. I do.

Indeed economists have written little about existential risk from AI. Of the work that has been done, most of it considers the tradeoff between AI-driven economic growth and existential risk. For example \citet{TrammellAschenbrenner2024} present a model in which accelerating technological development weakly reduces existential risk.\footnote{\citet{jones2016life} presents a model in which it is optimal to invest in safety as technology advances due to the increases in the value of life. For recent work on other margins of AI, see for instance \citet{motoki2024more, thompson2024ai, chugunova2025ruled}.} The reason is that accelerating technology can hasten the arrival of safety. \citet{jones2024ai} presents a model in which explosive growth due to AI only beats existential risk from the same when humans have near-log utility or anticipate large life-expectancy gains. \citet{Joneshowmuchshould2025} estimates that spending at least 1\% of GDP annually to avert catastrophic risk from AI is worthwhile. Such work takes for granted that misaligned ASI will harm humans. It does not consider the incentives that an misaligned ASI will have to help or harm humans.\footnote{\cite{gans2018self} is the rare exception. He argues that a paperclip-maximizing ASI may refrain from trying to seize all resources because it may not be able to control other AIs to whom it delegates the task of acquiring resources. As a result, a paperclip-maximizing ASI may be self-regulating. Unlike Gans, I consider a harder case in which the ASI is implicitly strong enough to control any sub-agents.} My article does. A related branch of work considers the problem of misaligned AI apart from existential risk. \citet{hadfield2016cooperative, hadfield2019incomplete, zhuang2020consequences}, for instance, aptly characterize the problem of misaligned AI as a principal-agent problem. They suggest that better contracts facilitate better behavior. By contrast I focus on the harder case in which AI is so strong that human principals do not have a third party capable of enforcing such contracts.

Lastly, I contribute to the economic analysis of non-human societies. Economists have shown that the economic approach can explain cooperation amongst ants and termites \citep{tullock1990economics, Tullock1992, tullock1994economics}, the allocation of labor effort among birds \citep{tullock1971coal}, externalities between plants and animals \citep{tullock1971biological}, and variations in animals' propensity to be malicious \citep{tullock1978altruism}. My paper suggests that the economic approach can predict the behavior of superintelligences just as it does low intelligences.\footnote{For related work in the human context, see for instance, \citet{becker1962irrational, gode1993allocative, leeson2022hobo}.}

\section{Context}
Agentic AI systems with general problem-solving abilities are likely to be created, deployed, and be acquisitive. Humans are likely to create such AI systems because they will be useful. The more tasks that can be delegated to such systems, the stronger humanity's incentive to create them. Labs like OpenAI and Google DeepMind have already declared plans to create artificial general intelligence (AGI). \citep{openai2025deepresearch, dragan2025responsible}.\footnote{Capable AI agents already exist. For example, organizations such as OpenAI have recently released an agentic coding tool called Codex that can achieve goals online. According to OpenAI, its employees have been able to delegate ``repetitive, well-scoped tasks" to Codex such as ``refactoring, renaming, and writing tests" and ``scaffolding new features, wiring components, fixing bugs, and drafting documentation" \citep{openai2025codex}. Anthropic's own agentic coding tool, Claude Code, ``delivers sustained performance on long-running tasks that require focused effort and thousands of steps, with the ability to work continuously for several hours" \citep{anthropic2025claude4}.} AGI refers to a computer system that is as capable as humans at any task. 

Moreover, highly capable, agentic AI systems are a real possibility. Consider first that AI technology is advancing very quickly \citep{EpochLLMBenchmarkingHub2024, epoch2023aitrends, AIIndex2025, macaskill2025preparing}. According to the Stanford Institute for Human-Centered AI's annual and comprehensive report on AI, high scores for the ``premier" general knowledge benchmark, Massive Multitask Language Understanding (MMLU), rose from 27.9\% in 2019 to 92.3\% in 2024, ``a remarkable 64.4 percentage point increase over five years" \citep[105]{AIIndex2025}.\footnote{AI is advancing rapidly along many other margins as well. For documentation of the rapid advances in visual reasoning, see \citet[121]{AIIndex2025}. For advances in speech recognition, see \citet[128]{AIIndex2025}. For advances in coding, see \citet[129]{AIIndex2025}. For advances in math, see \citet[134]{AIIndex2025}. For advances in reasoning, see \citet[138-141]{AIIndex2025}. For advances in AI agents, see \citet[148]{AIIndex2025} and for advances in robotics, see \citet[150]{AIIndex2025}.} 

Second, investment in AI is growing quickly. Data from \citet[248]{AIIndex2025} suggest that between 2013 and 2024, global private investment in AI has increased 2800\%, from \$5 billion to \$150 billion USD. In 2024 alone, AI attracted \$33.9 billion in global private investment, ``representing an 18.7\% increase from 2023 and over \textit{8.5 times the investment of 2022}" (italics added) \citep[248]{AIIndex2025}. Third, governments like the U.S. and China have a strong incentive to race to create highly advanced AI systems and are pouring billions of dollars into their creation \citep[4, 354]{AIIndex2025}.

Thanks to rapid progress in AI and the capital it has attracted, forecasters and AI researchers predict that transformative AI like AI will be created within the next century, if not sooner. As of June 4, 2025, Metaculus forecasters predict that the first general AI system will be devised, tested, and publicly announced in 2033. In 2020, the prediction was 2043.\footnote{Metaculus is an online, reputation-based prediction platform.} In 2024, \citet{grace2024thousands} asked 2,778 AI researchers to forecast when machines will outperform humans in every possible task. Respondents predicted that there was a 10\% chance that machines would outperform humans on every possible task by 2027 and a 50\% chance that machines would outperform humans by 2047.\footnote{For a more conservative prediction, see \citet{chow2023transformative}.}

Advanced AI systems will be goal-seeking by design. The general problem-solving abilities of an advanced AI system are part of what make it uniquely valuable. It is costly to find and then articulate to an AI the intended goal and the most efficient way in which to do a task.\footnote{Whether before, during, or after training or once it is deployed.} Specifying every possible state of the world and what to do in each would be prohibitively costly. AI's general problem-solving abilities allow humans to economize on such effort. Given a task, advanced AI can be left to find the best (and potentially novel) ways in which to execute it.\footnote{Analogs to this already exist amongst humans. It is often cheaper for a lawyer to give a mechanic a broad objective like ``my car is making an odd noise, please fix it" than it is to investigate the issue himself and then relay to the mechanic precisely how he wants it fixed.}

However, the same general problem-solving abilities that make an AI useful can make it harmful to humans. An autonomous AI system will not simply execute a task. It will reason about the most effective sub-goals for achieving its final objective. The problem is that such AI is very likely to choose subgoals that AI safety researchers consider dangerous. This is called the ``instrumental convergence thesis." That thesis says that AIs are likely to adopt subgoals such as resource acquisition, self-preservation, and self-improvement because they are very useful goals to have \citep{omohundro2018basic, bostrom2012superintelligent, turner2019optimal, soares2015corrigibility, benson2016formalizing, ji2023ai}. AIs with such subgoals are much more likely to be able to reach their ultimate goal.\footnote{Empirical work suggests that even rudimentary AI systems can pursue or show interest in such ``power-seeking" subgoals \citep{pan2023rewards, perez2023discovering, bondarenko2025demonstrating, meinke2024frontier}.} For example, an AI meant to clean a factory is far more likely to be able to accomplish that goal if it can amass the necessary inputs compared to one that does not accumulate resources.\footnote{Insofar as resource acquisition helps AI do more valuable economic tasks, humans may explicitly train models to adopt such subgoals.}

Such subgoals may be harmful because they may lead the AI to deviate from or impinge upon human interests. According to \citet[39]{hendrycks2023overview}, ``Power-seeking AIs with goals separate from ours are uniquely adversarial." The threat of power-seeking AI is twofold. First, AI may have the wrong objectives. The AI's ultimate reason for acquiring resources may be to pursue a goal that humans do not value. For example, an advanced AI system will produce goods that humans do not value. Second, an acquisitive AI may pursue its objectives in harmful ways. The AI will rationally choose the least-cost means of resource acquisition. For example, power-seeking AI will want property rights held by humans and may try to steal resources from them. 

The problem of misaligned AI is especially problematic if its abilities come to far exceed that of humans.\footnote{This may occur due to ``recursive self-improvement." According to \citet[29]{bostrom2014}, ``a process of recursive self-improvement might continue long enough to result in an intelligence explosion—an event in which, in a short period of time, a system's level of intelligence increases from a relatively modest endowment of cognitive capabilities . . . to radical superintelligence."} Such artificial superintelligence will also seek resources, power and self-preservation. But its enormous abilities mean that when it pursues the wrong goals or pursues goals in the wrong way, it will be unstoppable. \citet[123]{bostrom2014} has a famous example of a super capable but misaligned ``paperclip AI" that ``is given the final goal of maximizing the manufacture of paperclips, and proceeds by converting first the Earth and then increasingly large chunks of the observable universe into paperclips."

The extraordinary capabilities of artificial superintelligence may make it an existential risk to humans. \citet[115]{bostrom2014}, for example, suggests that ``a plausible default outcome of the creation of machine superintelligence is existential catastrophe." There are many ways in which ASI could pose an existential risk. ASI may create and deploy lethal bioweapons \citep{hendrycks2023overview}. \citet[4]{carlsmith2023existential} argues that ``If such agents are sufficiently capable . . . humans could end up permanently disempowered." Insofar as humans consist of useful resources or humans use resources that the ASI also wants, it will simply take them for its own \citep{bostrom2014}.

Because ASI is considered unstoppable, work in AI safety has focused on a technical solution: alignment. ``[T]he alignment problem," Dung states, ``is about building AI such that its goals conform to what its designers want it to do." \citep[138]{dung2023current}.\footnote{For similar descriptions, see \citet{ji2023ai}.} Examples include using reinforcement learning on human feedback to limit outer misalignment, reward modeling to reduce goal misgeneralization, and mechanistic interpretability to stop inner misalignment, adversarial training to improve AI alignment under distributional shifts, and quantilization to reduce reward hacking \citep{ji2023ai}. The hope is that ``[i]f we solve the alignment problem, then we avert this risk [of disempowerment]" \citep[138]{dung2023current}.

But even if humans do not want an AI to be misaligned, they may not be able to stop it. For example, a designer may fail to specify tolerable subgoals (or the final goal) for an AI to pursue. This is called the ``outer alignment" problem \citep{hubinger2019risks}. A designer can misspecify goals before training, during post-training, or after deployment. But even if the designer correctly specifies tolerable subgoals to an AI, an AI may learn during training to pursue subgoals that allow it to perform well during training but lead it to act against the designer's intent after deployment. This is the so-called ``inner alignment" problem \citep{hubinger2019risks}.\footnote{Inner misalignment can occur in various ways, including by reward hacking or through deceptive alignment.} The inner alignment problem is especially difficult to solve because designers cannot yet reliably detect when and why even rudimentary AI systems will pursue certain subgoals \citep[24]{sharkey2025open}, and advanced AIs are likely to have strong incentives to hide how aligned they are during training \citep{hubinger2019risks}. Without understanding an AI's inner reasoning, humans are unlikely to know until after deployment whether or not an AI is pursuing the right intermediate goals. In the case of ASI, that may be too late.

An advanced ASI is, therefore, likely to be misaligned. It will not consider the interests of humans. And because of its great abilities, humans have little chance of protecting themselves against it. 

\section{Theory}
It is tempting to think then that the extraordinary abilities of misaligned ASI will lead to catastrophe for humanity. But misalignment and vast disparities in strength are not enough to guarantee catastrophe.

To find the conditions under which humanity can avoid catastrophe despite enormous differences in power, I begin with a benchmark scenario in which humans can impose credible sanctions on ASI. I then progressively weaken humanity's position. I turn first to a world where the local ASI is too powerful to be sanctioned but it must compete with rivals for human output. Then I consider a world in which the local ASI is so strong that it has no rivals at all. I conclude with the bleakest scenario: a powerful ASI with no long-term stake in humanity's future. 

\subsection{Government sanctions}
Consider a misaligned, super-intelligent AI system. Misalignment means that the ASI assigns zero weight to human welfare. It wants to, for example, maximize paperclip output. That goal gives it a derived demand for zinc. The problem is that humans own the zinc.

The ASI can acquire the zinc in one of two ways. First, it can trade for zinc. Trading involves paying the market price in terms of cash, effort, or compute for a fixed quantity of zinc. Alternatively, the ASI can steal zinc. 

Here the ASI is not so strong that it can evade or avoid sanction. The humans are strong enough to impose sanctions via government but not strong enough to wrest all stealing capabilities from the ASI. Thus if the ASI steals then the humans can either do nothing or pay to sanction the ASI.

The ASI moves first and chooses whether to trade or steal. The humans then choose whether to sanction or not. If the ASI trades then the gains from trade are realized and the ASI earns $E_a$ and the humans earn their highest payoff $E_h$. The ASI buys zinc through the marketplace. 

If the ASI steals and the humans do nothing the ASI earns its highest payoff $S_a$ and the humans suffer harm $-H_h$ from losing zinc without compensation. However if the humans sanction the theft then the humans retake the zinc and punish the ASI for bad behavior. Sanctions may include, for example, fines, electricity and compute caps, or revoking internet access or any bonds the ASI may have posted for good behavior. When sanctioned the ASI suffers $ - P_a$ and the humans earn $-C_h$ from retaking the zinc and imposing the sanction. 

Because stealing avoids paying the market price for zinc, the ASI's payoff from stealing is greater than the payoff from trade, $S_a > E_a $. It has an incentive to steal. But because the sanction for stealing is harsh enough to be a deterrent, the ASI's payoff from trade is greater than the payoff from stealing and getting sanctioned, $E_a > - P_a$. To summarize, the ASI's payoffs are $S_a > E_a > - P_a$. 

From the humans' perspective, both inaction and sanctioning are costly. But the harm from stealing may be more or less than the costs of imposing sanctions, $-C_h \lesseqgtr -H_h$. In either case humans are better off when the ASI trades than when it steals, $E_h > -C_h \lesseqgtr -H_h$.

I solve the game via backwards induction. If the ASI steals then whether the humans sanction or not depends on whether the loss from imposing a sanction exceeds the loss from letting the ASI steal. Suppose that the harm from inaction is greater than the loss from imposing a sanction, $-C_h > -H_h$. In this case the humans will sanction the ASI if it steals. Anticipating the punishment the ASI will trade because the payoff from trading is greater than the payoff from stealing and then getting sanctioned, $E_a >- P_a$.

Now suppose that sanctioning the ASI is worse than the harm from inaction, $-C_h < -H_h$. In this case the humans will not sanction the ASI if it steals. The ASI will therefore always steal because the payoff from stealing is greater than the payoff from trading, $S_a > E_a$.

As a result there are two possible outcomes. If $-C_h > -H_h$, then the humans' sanction is credible. The ASI trades to avoid the sanction and the humans do nothing, yielding payoffs $(E_a, E_h)$. But if $-C_h < -H_h$ then the humans' sanction is not credible. The ASI steals and the humans do nothing and the payoffs are $(S_a, -H_h)$. 

The foregoing analysis offers three insights. First, the ASI will only harm humans when sanctions are not credible. Harm requires that $-C_h < -H_h$. If, for example, the harms from stealing are low or the costs of imposing sanctions are high, then humans will not find sanctions worthwhile. The optimal quantity of ASI harms may indeed be positive. Just as societies tolerate some crime because the costs of perfect enforcement would be prohibitive, humanity may tolerate some transgressions from ASI. 

Second, even a misaligned ASI will respond to incentives. Even though the ASI may have a terminal goal we do not know or understand, it will economize. This means that the ASI will choose the least-cost means with which to pursue its final goal, whatever it may be. The reason is that economizing will give the ASI the best chance at reaching its final goal. This has two implications. Like self-preservation and acquisitiveness, ``economizing" may also be an instrumentally convergent subgoal. And since credible sanctions make trade the least-cost method with which it can acquire resources, even a misaligned ASI will follow the law. 

Three, the logic of credible sanctions extends beyond simple goods like zinc. It holds whether the ASI $i)$ has a derived demand for a narrow subset of goods like chips, $ii)$ wants property rights to political power, human labor, or raw materials or $iii)$ can acquire resources through force, fraud, or blackmail. In any such case the threat of credible sanctions may be sufficient to induce cooperation from a moderately powerful ASI.\footnote{The model also applies to situations in which the ASI does not directly steal resources from humans but instead produces negative externalities. Pollution may, for example, be a byproduct of the ASI making paperclips. Pollution reduces the value of resources used by humans without paying them for that loss. This is analogous to stealing the environmental resource from humans. Sanctions can then discourage pollution.}

\subsection{Competition}
In the benchmark model the ASI's cooperation is compelled by a third party: the government. This rosy outcome hinges on the assumption that the government is strong enough to impose credible sanctions. But sanctions do not work when the ASI is far stronger than any human government. Any attempt to punish the ASI it can unwind or evade. This is the canonical misaligned ASI. To capture this more challenging scenario, I relax the assumption that humans can credibly sanction the ASI.

As the benchmark model suggests, such an all-powerful ASI will always steal from humans in a one-shot interaction because stealing avoids having to pay for zinc, $S_a > E_a$. Humanity's situation looks dire indeed. It is therefore tempting to conclude that without government capable of checking ASI, unrestrained predation would occur. But that may not always happen.

An extraordinarily strong ASI can take at will from humans. But consider a world in which there are other, equally strong ASIs nearby. A local ASI may be able to take from humans but it cannot do the same to its rivals. Ergo each ASI has a monopoly on theft in its respective area. 

The prospect of rivals gives humans and the local ASI each a new strategic choice. If the local ASI steals, humans now have the option to flee. Fleeing involves moving part of humans' zinc capacity from the area of the local ASI to that of a rival's. The local ASI, in turn, now has the option to skim. Skimming is less punitive than stealing and leaves enough zinc to discourage human exit.

The game is now infinitely repeated. Each period, the ASI can trade, steal, or skim. Humanity observes the ASI's choice and then decides whether to do nothing, sanction, or flee. If the ASI trades then for that period it earns the low payoff of $E_a$ and the humans earn their highest payoff $E_h$.

If the local ASI steals and the humans pick inaction then the local ASI earns $S_a$ and the humans earn $-H_h$ for that period. If instead the humans sanction the ASI it still earns $S_a$ and the humans lose both the value of the zinc $-H_h$ and the costs of the sanction $-C_h$ since the ASI is immune to sanction. Lastly, humans can flee. If some humans flee to a rival ASI they lose $-H_h$ in that period. However, in all subsequent periods they lose only $-f_h$ since the rival steals less.\footnote{A rival would steal less when, for example, it has a low demand for zinc, is more benevolent, or has a comparative advantage in stealing.} As a result the local ASI earns $S_a$ in the first period and then earns only $s_a$ in every later period since humans take their productive capacity with them.

But the ASI need not steal all the zinc. It can also skim. If the ASI skims and the humans choose inaction, the local ASI earns less from stealing, $T_a$. The humans then lose only $-h_h$. Should the humans try to resist and sanction the ASI then they earn $-h_h - C_h$ and the ASI still earns $T_a$ because sanctions are ineffective. As above if some humans flee they again earn $-h_h$ in the first period and $-f_h$ thereafter. The local ASI in turn earns $T_a$ in the first period and then $s_a$ in every other period.

To summarize, the ASI's payoffs are $S_a > T_a > s_a > E_a$. The ASI earns the most from stealing, somewhat less from skimming, even less when the humans flee, and the least when it decides to trade. By contrast the humans' payoffs are $E_h >  -h_h > -f_h  >-h_h - C_h >-H_h$. Humanity earns the most when the ASI trades, somewhat less when their zinc is skimmed, even less when they flee, and the least when they pick inaction when the ASI steals. 

I assume that humans play a grim trigger strategy. Should the ASI ever steal, the humans flee forever.\footnote{The threat of permanent flight is credible because humanity's present discounted value of flight is greater than the present discounted value of staying and doing nothing. The Humans flee when
\begin{equation*}
    -H_h  +   \frac{-f_h \delta_h}{1 - \delta_h} \geq \frac{-H_h}{1 - \delta_h}
\end{equation*}
or
\begin{equation*}
    H_h \geq f_h.
\end{equation*}
Because the harm from stealing is greater than the harm from flight, the threat of exit is credible.} Humans never sanction because it has no benefits and reduces humanity's payoff. As a result, if the ASI steals, the humans' best response is to flee because $-f_h > -H_h $. If the ASI skims, the humans' best response is to do nothing because $-h_h> -f_h $. 

The ASI anticipates flight if it steals. It never trades because the payoff from trade is always dominated by stealing or skimming. As a result the ASI weighs the payoff from skimming against that of stealing. The present discounted value of stealing is
\begin{equation*}
S_a  +   \frac{s_a \delta_a}{1 - \delta_a},
\end{equation*}
where $\delta_a \in (0,1)$ is the ASI's discount factor. The present discounted value of skimming is 
\begin{equation}
    \frac{T_a}{1 - \delta_a}.
\end{equation}
The ASI skims rather than steals as long as 

\begin{equation*}
      \frac{T_a}{1 - \delta_a} \geq S_a  +   \frac{s_a \delta_a}{1 - \delta_a}
\end{equation*}
or
\begin{equation}
    \delta_a (S_a - s_a) \geq  S_a - T_a
\end{equation}
This expression says that the ASI will choose to skim rather than steal when the discounted value of lost output ($\delta_a (S_a - s_a)$) is at least as large as the immediate gain from stealing today ($S_a - T_a$).

There are two possible outcomes. If the local ASI is sufficiently patient then it skims and the humans stay, yielding payoffs $(T_a, -h_h)$ every period. But if the local ASI is impatient then it steals in the first period and then humans flee, yielding $(S_a, -H_h)$ in the first period. In every period thereafter the payoffs change to $(s_a, -f_h)$ since ASI steals less and the humans are skimmed by the rival ASI.

This extension offers two insights. First, sanctioning is never rational in the presence of canonical ASI. Because the ASI is so strong, any attempt to sanction the ASI is not worthwhile. Second, mobility and competition can help humans avoid catastrophe despite the local ASI's enormous strength. As long as humans have an exit option, a ``market for theft" will emerge in which extraordinarily strong ASIs compete by offering, for example, to take less from humans. To the great benefit of humanity, an extraordinarily strong ASI will not necessarily engage in unrestrained banditry. 

\subsection{Encompassing interest}
But humans may have no exit option. This may happen when, for example, they cannot flee, the ASIs collude with one another, or there is one ASI whose strength far exceeds that of its rivals. In this case, humans are subject to predation by a true monopolist. Although human welfare is worse under such an ASI, it is still likely that they will avoid catastrophe.

The reason is that the ASI's dominance ties its wealth to the long-run productivity and prosperity of humans \citep{olson1993dictatorship}. Stealing too much today depresses human productivity and humanity's incentive to make zinc tomorrow.\footnote{Even if the ASI is stronger than humans it is likely to value their output due to humanity's comparative advantage. Suppose the ASI has an absolute advantage in both strength and in making zinc. However, the opportunity cost of using cognitive resources to do mundane tasks could be quite large. That same processing power might be better used working on interstellar engineering problems or designing nanomachines. Humans, by contrast, have a lower opportunity cost for producing zinc. It is therefore cheaper for the ASI to allow humans to specialize in tasks where they have a comparative advantage, freeing up its own resources for tasks only it can perform. This creates the very tax base that the ASI has an encompassing interest in preserving.} As a result a truly dominant ASI has a \textit{de facto} but genuine property right to human output. To maximize the present discounted value of that property right, such an ASI will weigh the short-run gains from stealing a lot today but less tomorrow against stealing moderate amounts of zinc today and tomorrow.  

The prospect of humans making less zinc gives the ASI a new strategic choice in place of skimming: taxation. Taxing is a kind of theft. Humans get nothing in return. However, taxation leaves just enough zinc to humans that they have an incentive to keep making zinc in the future.

As before, the game is infinitely repeated. Each period, the ASI can now trade, steal, or tax. Because sanctioning is never rational and humans can no longer flee, humans have no choices. They are at the whim of ASI.

If the ASI trades the ASI earns the low payoff of $E_a$ and the humans earn their highest payoff $E_h$. If the ASI steals the ASI earns $S_a$ and the humans get $-H_h$. If the ASI taxes and the humans do nothing the ASI earns a moderate payoff of $T_a$. The humans lose only $-h_h$. 

As before the payoff from stealing falls permanently to $s_a$ in subsequent periods but for a different reason. After losing everything, humans no longer have an incentive to make as much zinc in the future.\footnote{The payoff from stealing could also fall because it permanently destroys human productive capacity in terms of labor and capital.} Because the ASI takes all that remains in subsequent periods, humanity's loss is normalized to $-H_h$ in future periods. By contrast the payoff from trade and taxation stays the same in every period. The payoff from taxation stays the same every period because the ASI leaves enough zinc for humans that they have an incentive to keep making zinc. Because taxation takes less than stealing but does not involve paying humans, $S_a > T_a > s_a > E_a $. The humans' payoffs are $E_h > -h_h > - H_h $. Humans benefit the most from trade and lose more from stealing than they do from taxation.

Because humans are at the whim of ASI, payoffs only depend upon the actions of ASI. The ASI is free to choose among trading, taxing, or stealing. To do so the ASI weighs the present discounted value of always trading to always taxing to always stealing, where $\delta \in (0,1)$ is the ASI's discount factor. The present discounted value of always trading is
\begin{equation*}
    \frac{E_a}{1-\delta}.
\end{equation*}

The present discounted value of always taxing is
\begin{equation*}
    \frac{T_a}{1-\delta}.
\end{equation*}
Because the payoff from stealing falls to $s_a$ in all future periods when the ASI steals in the first period, the present discounted value of always stealing is
\begin{equation*}
    S_a + \frac{\delta s_a}{1-\delta}.
\end{equation*}

The ASI never trades. As the one-shot payoff from taxation exceeds the one-shot payoff from trade, the present discounted payoff from taxation is also larger than that of trading. 

However, the ASI will choose to tax rather than steal as long as 
\begin{equation*}
        \frac{T_a}{1-\delta} \geq  S_a + \frac{\delta s_a}{1-\delta}
\end{equation*}
or
\begin{equation}
    \delta (S_a - s_a) \geq  S_a - T_a
\end{equation}
As above, this expression says that the ASI will choose to tax rather than steal when the discounted value of lost output ($\delta (S_a - s_a)$) is at least as large as the immediate gain from stealing today ($S_a - T_a$).

Because the ASI never trades, there are two possible outcomes neither of which depends upon the humans' actions. If the ASI is sufficiently patient then it taxes and the humans pick inaction, yielding payoffs $(T_a, -h_h)$ each period. If the ASI is impatient, the payoffs are $(S_a, -H_h)$ in the first period and $(s_a,-H_h)$ thereafter.

The foregoing analysis yields four additional insights. First, an extraordinarily powerful ASI is unlikely to constantly steal from humans. In this extension maximal harm requires that the ASI be impatient, $\delta <  \frac{S_a - T_a}{S_a - s_a}$. But a future ASI is likely to be quite patient. Humans are mortal. By contrast a digital ASI has no natural lifespan as it can be backed up. For an entity with such a long lifespan, the vast majority of its existence is in the future, compelling it to weigh future outcomes far more heavily than a human would. As a result a superhuman ASI is likely to be sufficiently patient that it will tax humans rather than steal from them.

Second, the model suggests that an extraordinarily powerful ASI will not be a chaotic ravager. It will be a rational autocrat. As a result human welfare may not be as low as one might think when ASI is extraordinarily strong. Of course compared to a world in which humans can impose credible sanctions on ASI, humans are worse off. But an ASI's encompassing interest checks its incentive to hurt humanity.\footnote{As above, this extension also applies to situations in which the ASI produces negative externalities. Rather than stealing or taxation, in this case the ASI chooses between different levels of pollution: a lot or a little. If sufficiently patient then the ASI will, to the humans' delight, choose a sustainable level of pollution.} Hurting humans means that it also hurts its own long-run interests. Indeed, ASI may have an incentive to nurture humanity because humans generate benefits for it. An encompassing interest and long time horizons put an upper bound on just how bad it will be for humans.

Third, attempts to design AIs with short-term or ``myopic" preferences may backfire. \citet{carlsmith2022power, thornley2024towards} and \citet{thornley2025shutdownable} suggest that AIs with short-term interests are less likely to, for example, seek power or resist shutdown. My analysis suggests that such attempts at alignment risk backfiring because they encourage powerful AI to acquire resources in particularly harmful ways.

Fourth, even a misaligned ASI might choose to tie its own hands. Rulers sometimes limit their own discretion to get subjects to invest more \citep{barzel1997development, barzel1997parliament, barzel2000property}. The same logic applies here. By committing to clear limits such as a fixed tax schedule, due process before taking, or a hard cap on its share, the ASI may be able to induce humans to make even more zinc. Doing so is costly. It reduces the ASI's ability to grab everything today. However, credible commitments to not grab may give the ASI a larger base from which to skim tomorrow. If the extra zinc generated by tying its own hands is big enough to outweigh the cost of the constraint, the ASI wins in present value. 

\subsection{Trading on credit}
In the previous extension, an encompassing interest checks an all-powerful ASI's incentive to prey upon humans. I now relax that assumption to analyze the most challenging scenario yet: humans face an extraordinarily strong ASI that only gets one opportunity to grab from humans. This could happen when, for example, there are many superhuman AIs, each with a tenuous claim to zinc held by humans. Without the shadow of the future to create an encompassing interest, one might expect full predation. But this need not always be the case.\footnote{The model below draws heavily from \citet{leeson2007trading}.}

The game is now one-shot. Humans can either produce zinc for trade or produce zinc for subsistence. Producing zinc for trade involves creating a great deal of zinc. Producing zinc for subsistence entails producing only enough zinc for self-consumption. 

Now the humans move first.\footnote{The ASI cannot steal zinc that does not exist. Therefore the humans must move first.} If the humans produce for trade then the ASI sees a large surplus of zinc ripe for taking. It chooses whether to trade, steal, or ignore humans.\footnote{Because the game is one-shot, taxing is dominated by stealing. As a result, I drop that option from the ASI's set of choices.} If the ASI steals, then the ASI takes all the zinc and the game ends. The ASI gets its biggest payoff $S_a$ and the humans get their lowest payoff $-H_h$. If the ASI trades, then the ASI buys zinc from the humans and the game ends. The ASI gets its second-best payoff $E_a$ and the humans get their highest possible payoff $E_h$. If the ASI ignores the humans, the ASI gets its third-best payoff $I_a$. Since humans produce a great deal of zinc and are left alone, humans get their second-best payoff $I_h$.

If the humans produce for subsistence then the ASI can again steal, trade, or ignore. If the ASI steals, it takes what little the humans have and the game ends. The ASI gets its smallest payoff $s_a$ and the humans lose only $-h_h$. If the ASI ignores the humans, it still gets its third-best payoff $I_a$. By contrast humans get a low payoff of $i_h$ since they are only producing enough zinc to consume for themselves.

But if the ASI decides to trade when the humans produce for subsistence, trade occurs over two subperiods. In subperiod 1 the ASI pays in advance and humans promise to make zinc in subperiod 2. In subperiod 2, if the humans carry out their promise then the humans pay what they owe to the ASI and get the value from selling zinc $E_h$, and the game ends. ASI gets the present discounted value buying zinc, $\beta E_a$ where $\beta \in (0,1)$ is the discount factor for each subperiod. However, if the humans break their promise and make nothing then the ASI takes as much as it can to recoup the debt in subperiod 2 and the game ends. The ASI's payoff is then $\beta s_a$, the discounted value of stealing what little the humans have left. The humans' payoff is the advance they get in subperiod 1 less the discounted value of what is taken in subperiod 2, $E_h - \beta h_h$.

The ASI's payoffs are therefore ranked as follows, $S_a > E_a > I_a > s_a$.\footnote{The assumption $I_a > s_a$ is realistic as long as the ASI stealing has an opportunity cost. Because humans produce so little zinc, any energy, compute, or attention used to steal will be better used elsewhere. Ignoring the humans is then better than stealing next to nothing.} The ASI gets the most from stealing when the humans produce for trade, even less from trade, even less from ignoring the humans, and the least from stealing when the humans produce for subsistence. The humans' payoffs are $E_h > I_h > i_h > - h_h > -H_h$. Humans get the most from trade, even less from being ignored when they produce for trade, and even less when they are ignored and produce for subsistence. Humans' payoffs go negative when they are stolen from but they lose more from theft when they produce for trade than when they produce for subsistence.

The humans never break their promise if the ASI pays for zinc in advance. Whether the ASI steals, trades, or ignores the humans depends upon its discount rate. If the humans produce for subsistence and $\beta \geq \frac{I_a}{E_a}$, the ASI pays on credit. Otherwise, it ignores the humans because $I_a > s_a$. If the humans produce for trade, the ASI always steals because $S_a > E_a > I_a$. 

The humans know that if they produce for trade then the ASI will steal and they will earn their lowest payoff $-H_h$. But if they produce for subsistence they get $i_h$ if the ASI is insufficiently patient and $E_h$ if the ASI is sufficiently patient. As a result there are but two possible outcomes. 

If $\beta \geq \frac{I_a}{E_a}$ then the humans produce for subsistence, the ASI pays in advance, the humans perform as promised and the pair trade, yielding payoffs $(E_a, E_h)$. If $\beta < \frac{I_a}{E_a}$ humans produce for subsistence and the ASI ignores the humans yielding payoffs $(I_a, i_h)$.

This last extension yields two final insights. First, a one-shot game with an extraordinarily strong ASI does not guarantee theft. To the contrary, humans can induce cooperation from the ASI by threatening not to make anything. The ASI cannot take that which does not yet exist. So even though the humans cannot sanction the ASI \textit{ex post}, they can punish the ASI \textit{ex ante}. By withholding future output, humans can incentivize the ASI to bribe them today. And as long as the ASI is sufficiently patient, it will pay.

Relatedly, a strong ASI without an encompassing interest does not necessarily lead to the worst imaginable outcome for humans. Compared to a world in which the ASI has an encompassing interest, humans are worse off. But even in a one-shot game the worst-case scenario of full predation is avoided. Humans' ability to withhold production of zinc checks the ASI's incentive to steal.

\section{Conclusion}
What if the ASI enslaves humans? Enslavement fits naturally into the foregoing discussion of an ASI with an encompassing interest. In economic terms, enslavement refers to an owner who holds a property right to a subject's output but does not pay market compensation for that output. In the case of ASI, it becomes a rational slave-owner rather than a rational autocrat. As a result the same logic that guides the ASI to leave humans with enough resources to remain productive also applies here. And insofar as torture reduces human output, a patient ASI would give, for example, healthcare, sustenance, and education to its human assets if doing so maximized the long-term stream of extracted value. Thus, even in the grim scenario of enslavement, the ASI's own self-interest provides a shield against gratuitous cruelty.

My analysis focuses on an acquisitive ASI, but the framework also sheds light on harms caused by an ASI that is indifferent to resources held by humans. This may occur when, for example, the ASI is self-sufficient and does not rely on human inputs. Such an agent would have little reason to steal. Nevertheless it might pursue its goals in ways that impose large external costs on humanity. In this case humanity may be able to bribe ASI to abate its harmful activities if the costs of defining and enforcing the terms of that bribe are low enough \citep{coase1960problem}. The logic holds even with large-scale harms. The reason is that the larger the external costs, the greater is humanity's incentive to find a way to pay the bribe. Even an indifferent superintelligence may, therefore, have an incentive to treat humans well.

My analysis yields three conclusions. First, misalignment and awesome abilities are not sufficient for an ASI to take part in unrestrained theft. Existential-style predation requires simultaneous failure of enforcement, exit, and patience. If any one margin is strong enough, full confiscation is not an equilibrium. While humanity's welfare is highest when it can enforce its own rules, interjurisdictional competition and patience can provide a surprising degree of protection against the worst possible outcomes.

Second, my last extension suggests that cooperation with a super-capable agent does not require alignment, equal strength, repeated dealings, or third-party enforcement. It is sufficient that an ASI be somewhat patient. In that case ASI will have little incentive to prey upon humanity since it cannot seize value that has not yet been created. Humanity's ability to withhold future output therefore becomes a bargaining chip. Without being paid in advance, humans will produce so little that predation is not worthwhile. As long as the ASI is somewhat patient, trade will occur. Naturally human welfare is far from ideal in this world. But the prospect of cooperation even in such dire circumstances ought to give economists pause. Should cooperation be possible here, then cooperation also ought to be possible in a wide range of other, equally dire situations. 

Lastly, the very rapaciousness that makes ASI a threat to humans also disciplines its behavior. An acquisitive ASI that engages in unrestrained predation destroys a key source of wealth: human output. As a result the ASI's own self-interest tempers its short-term rapaciousness. The ASI will rule as a rational autocrat. As a little-known economist once wrote, ``[i]t is not from the benevolence of the butcher, the brewer, or the baker that we expect our dinner, but from their regard to their own interest. We address ourselves, not to their humanity but to their self-love," \citep[26-27]{smith_wealth_1981}. The same, I contend, is true of alien intelligences. Just as Smith showed how self-interest could lead to mutual prosperity, my paper suggests that self-love may also limit ASI's poor treatment of humans.

\newpage
\bibliographystyle{apalike}
\bibliography{References}

\end{document}